\documentclass[11pt]{article} 

\usepackage{a4wide}

\usepackage{fullpage}
\usepackage{amssymb} \usepackage{amsmath} \usepackage{bm}
\usepackage{times} \usepackage{fancybox} \usepackage{color}
\usepackage{latexsym}
\usepackage{eepic} \usepackage{epic} \usepackage{epsf}
\usepackage{graphics}  \usepackage{pstricks}
\psset{unit=.5pt} \newtheorem{theorem}{Theorem}[section]
\newtheorem{lemma}[theorem]{Lemma}

\newtheorem{corollary}[theorem]{Corollary}

\newtheorem{observation}[theorem]{Observation}

\newtheorem{remark}[theorem]{Remark}  \newcommand{\bproof}{\noindent{\it Proof}}
\newcommand{\cproof}{\noindent{\it Proof of Claim}}
\newcommand{\eproof}{\hspace*{\fill}$\rule{2mm}{2mm}$~~~~~\bigskip}
\renewenvironment{proof}{\bproof. }{\eproof}

\newcommand{\PIT}{\mbox{\small\rm PIT}}

\newcommand{\Perm}{\ensuremath{Perm}}

\newcommand{\NEXP}{\mbox{\small\rm NEXP}}

    \newcommand{\MA}{\mbox{\rm MA}}

  \renewcommand{\P}{\mbox{\rm P}}

\newcommand{\ppoly}{\mbox{\rm P/poly}}
\newcommand{\NSUBEXP}{\mbox{\rm NSUBEXP}}
\newcommand{\SUBEXP}{\mbox{\rm SUBEXP}}
\newcommand{\DTIME}{\ensuremath{\rm DTIME}}
\newcommand{\NTIME}{\ensuremath{\rm NTIME}}

\newcommand{\NL}{\mbox{\rm NL}} 

\newcommand{\UL}{\mbox{\rm UL}}

\newcommand{\RNC}{\ensuremath{\rm RNC}} \newcommand{\poly}{\mbox{\rm
poly}}

\newcommand{\size}{\ensuremath{\it size}}

\newcommand{\Q}{\mathbb{Q}} \newcommand{\Z}{\mathbb{Z}}
 
\newcommand{\F}{\ensuremath{\mathbb{F}}}

 \newcommand{\Prob}{\mbox{\rm Prob}}

\newcommand{\A}{\ensuremath{\mathcal{A}}}

\title{Derandomizing the Isolation Lemma and Lower Bounds for
  Circuit Size}

\author{V.~Arvind and Partha Mukhopadhyay\\
Institute of Mathematical Sciences\\ C.I.T Campus,Chennai  600 113,
India\\
\tt{\{arvind,partham\}@imsc.res.in}
}

\date{}

\begin{document}

\maketitle

\begin{abstract}
  The isolation lemma of Mulmuley et al \cite{MVV87} is an important
  tool in the design of randomized algorithms and has played an
  important role in several nontrivial complexity upper bounds.  On
  the other hand, polynomial identity testing is a well-studied
  algorithmic problem with efficient randomized algorithms and the
  problem of obtaining efficient \emph{deterministic} identity tests
  has received a lot of attention recently. The goal of this note is
  to compare the isolation lemma with polynomial identity testing:
\begin{enumerate}
\item We show that derandomizing reasonably restricted versions of the
  isolation lemma implies circuit size lower bounds. We derive the
  circuit lower bounds by examining the connection between the
  isolation lemma and polynomial identity testing. We give a
  randomized polynomial-time identity test for noncommutative circuits
  of polynomial degree based on the isolation lemma. Using this
  result, we show that derandomizing the isolation lemma implies
  noncommutative circuit size lower bounds. The restricted versions of
  the isolation lemma we consider are natural and would suffice for
  the standard applications of the isolation lemma.

\item From the result of Klivans-Spielman \cite{KS01} we observe that
  there is a randomized polynomial-time identity test for commutative
  circuits of polynomial degree, also based on a more general
  isolation lemma for linear forms. Consequently, derandomization of
  (a suitable version of) this isolation lemma implies that either
  $\NEXP\not\subset \P/\poly$ or the Permanent over $\Z$ does not have
  polynomial-size arithmetic circuits.
\end{enumerate}

\end{abstract}

\section{Introduction}\label{intro}\label{isolation}

We recall the Isolation Lemma \cite{MVV87}. Let $[n]$ denote the set
$\{1,2,\cdots,n\}$. Let $U$ be a set of size $n$ and
$\mathcal{F}\subseteq 2^U$ be any family of subsets of $U$. Let $w :
U\rightarrow\mathbb{Z}^{+}$ be a weight function that assigns positive
integer weights to the elements of $U$. For $T\subseteq U$, define its
weight $w(T)$ as $w(T)=\sum_{u\in T} w(u)$. Then Isolation Lemma
guarantees that for any family of subsets $\mathcal{F}$ of $U$ and for
any random weight assignment $w : U\rightarrow [2n]$, with high
probability there will be a unique minimum weight set in
$\mathcal{F}$.

\begin{lemma}[Isolation Lemma]\rm{\cite{MVV87}}\label{iso-lemma}
  Let $U$ be an universe of size $n$ and $\mathcal{F}$ be any family
  of subsets of $U$. Let $w : U\rightarrow [2n]$ denote a weight
  assignment function to elements of $U$. Then,
\[
\Prob_w[~\mbox{There exists a unique minimum weight set in $\mathcal{F}$}]
\geq\frac{1}{2}, 
\]
where the weight function $w$ is picked uniformly at random.
\end{lemma}

In the seminal paper \cite{MVV87} Mulmuley et al apply the isolation
lemma to give a randomized NC algorithm for computing maximum
cardinality matchings for general graphs (also see \cite{ARZ}). Since
then the isolation lemma has found several other applications. For
example, it is crucially used in the proof of the result that
$\NL\subset \UL/\poly$ \cite{AR} and in designing randomized NC
algorithms for linear representable matroid problems \cite{NSV90}. It
is also known that the isolation lemma can be used to prove the
Valiant-Vazirani lemma that SAT is many-one reducible via randomized
reductions to USAT.

Whether the matching problem is in deterministic NC, and whether
$\NL\subseteq\UL$ are outstanding open problems. Thus, the question
whether the isolation lemma can be derandomized is clearly important.

As noted in \cite{Agr07}, it is easy to see by a counting argument
that the isolation lemma can not be derandomized, in general, because
there are $2^{2^n}$ set systems $\mathcal{F}$. More formally, the
following is observed in \cite{Agr07}.

\begin{observation}{\rm{\cite{Agr07}}}
  The Isolation Lemma can not be fully derandomized if we allow weight
  functions $w:U\rightarrow [n^c]$ for a constant $c$ (i.e.\ weight
  functions with a polynomial range). More precisely, for any
  polynomially bounded collection of weight assignments
  $\{w_i\}_{i\in[n^{c_1}]}$ with weight range $[n^{c}]$, there exists
  a family $\mathcal{F}$ of $[n]$ such that for all $j\in[n^{c_1}]$,
  there exists two minimal weight subsets with respect to $w_j$.
\end{observation}

However that does not rule out the derandomization of any special
usage of the isolation lemma. Indeed, for all applications of the
isolation lemma (mentioned above, for instance) we are interested only
in exponentially many set systems $\mathcal{F}\subseteq 2^U$. 

We make the setting more precise by giving a general framework. Fix
the universe $U=[n]$ and consider an $n$-input boolean circuit $C$
where $\size(C)=m$. The set $2^U$ of all subsets of $U$ is in a
natural $1$-$1$ correspondence with the length $n$-binary strings
$\{0,1\}^n$: each subset $S\subseteq U$ corresponds to its
characteristic binary string $\chi_S\in\{0,1\}^n$ whose $i^{th}$ bit
is $1$ iff $i\in S$.  Thus the $n$-input boolean circuit $C$
implicitly defines the set system
\[
\mathcal{F}_C=\{S\subseteq[n]\mid C(\chi_S)=1\}.
\]

As an easy consequence of Lemma~\ref{iso-lemma} we have
the following.

\begin{lemma}\label{mn-iso-lemma}
  Let $U$ be an universe of size $n$ and $C$ be an $n$-input boolean
  circuit of size $m$. Let $\mathcal{F}_C\subseteq 2^U$ be the family
  of subsets of $U$ defined by circuit $C$. Let $w : U\rightarrow
  [2n]$ denote a weight assignment function to elements of $U$. Then,
\[
\Prob_w[~\mbox{There exists a unique minimum weight set in $\mathcal{F}_C$}]
\geq\frac{1}{2}, 
\]
where the weight function $w$ is picked uniformly at random.
Furthermore, there is a collection of weight functions $\{w_i\}_{1\leq
  i\leq p(m,n)}$, where $p(m,n)$ is a fixed polynomial, such that for
each $\mathcal{F}_C$ there is a weight function $w_i$ w.r.t.\ which
there is a unique minimum weight set in $\mathcal{F}_C$.
\end{lemma}

Lemma~\ref{mn-iso-lemma} allows us to formulate two natural and
reasonable derandomization hypotheses for the isolation lemma.\\

\noindent\textbf{Hypothesis 1.}~~ There is a deterministic algorithm
$\mathcal{A}_1$ that takes as input $(C,n)$, where $C$ is an $n$-input
boolean circuit, and outputs a collection of weight functions
$w_1,w_2,\cdots,w_t$ such that $w_i:[n]\rightarrow [2n]$, with the
property that for some $w_i$ there is a unique minimum weight set in
the set system $\mathcal{F}_C$. Furthermore, $\mathcal{A}_1$ runs in
time subexponential in $\size(C)$.

\noindent\textbf{Hypothesis 2.}~~ There is a deterministic algorithm
$\mathcal{A}_2$ that takes as input $(m,n)$ in unary and outputs a
collection of weight functions $w_1,w_2,\cdots,w_t$ such that
$w_i:[n]\rightarrow [2n]$, with the property that for each size $m$
boolean circuit $C$ with $n$ inputs there is some weight function
$w_i$ w.r.t.\ which $\mathcal{F}_C$ has a unique minimum weight set.
Furthermore, $\mathcal{A}_2$ runs in time polynomial in $m$.\\

Clearly, Hypothesis 2 is stronger than Hypothesis 1. It demands a
``black-box'' derandomization in the sense that $\mathcal{A}_2$
efficiently computes a collection of weight functions that will work
for \emph{any} set system in $2^U$ specified by a boolean circuit of
size $m$.

Notice that a random collection $w_1,\cdots,w_t$ of weight functions
will fulfil the required property of either hypotheses with high
probability. Thus, the derandomization hypotheses are plausible.
Indeed, it is not hard to see that suitable standard hardness
assumptions that yield pseudorandom generators for derandomizing BPP
would imply these hypotheses. We do not elaborate on this here. In
this paper we show the following consequences of Hypotheses 1 and 2.

\begin{enumerate}
\item Hypothesis 1 implies that either $\NEXP\not\subset \P/\poly$ or
  the Permanent does not have polynomial size noncommutative
  arithmetic circuits.
\item Hypothesis 2 implies that for each $n$ there is an explicit
  polynomial $f_n(x_1,x_2,\cdots,x_n)\in\F\{x_1,x_2,\cdots,x_n\}$ in
  noncommuting variables $x_i$ (where by explicit we mean that the
  coefficients of the polynomial $f_n$ are computable by a uniform
  algorithm in time exponential in $n$) that does not have
  noncommutative arithmetic circuits of size $2^{o(n\lg n)}$ (where
  the field $\F$ is either the rationals or a finite field).
\end{enumerate}

These two results are a consequence of an identity testing algorithm
for noncommutative circuits that is based on the isolation lemma.
This algorithm is based on ideas from \cite{AMS08} where we used
automata theory to pick matrices from a suitable matrix ring and
evaluate the given arithmetic circuit on these matrices. In the next
section, we describe the background and then give the identity test in
the following section.

\begin{remark}
  Notice that derandomizing the isolation lemma in specific
  applications like the $\RNC$ algorithm for matchings
  {\rm\cite{MVV87}} and the containment $\NL\subseteq\UL/\poly$
  {\rm\cite{AR}} might still be possible without implying such circuit
  size lower bounds.
\end{remark}

Noncommutative polynomial identity testing has been the focus of
recent research \cite{RS05,BW05,AMS08}. One reason to believe that it
could be easier than the commutative case to derandomize is because
lower bounds are somewhat easier to prove in the noncommutative
setting as shown by Nisan \cite{N91}. Using a rank argument Nisan has
shown exponential size lower bounds for noncommutative formulas (and
noncommutative algebraic branching programs) that compute the
noncommutative permanent or determinant polynomials in the ring
$\F\{x_1,\cdots,x_n\}$ where $x_i$ are noncommuting variables.
However, no superpolynomial lower bounds are known for the size of
noncommutative circuits for explicit polynomials.

Our result in this paper is similar in flavor to the
Impagliazzo-Kabanets result \cite{KI03}, where for \emph{commutative}
polynomial identity testing they show that derandomizing polynomial
identity testing implies circuit lower bounds. Specifically, it
implies that either $\NEXP\not\subset \ppoly$ or the integer Permanent
does not have polynomial-size arithmetic circuits.

In \cite{AMS08} we have observed that an analogous result also holds
in the noncommutative setting.  I.e., if noncommutative PIT has a
deterministic polynomial-time algorithm then either
$\NEXP\not\subset \ppoly$ or the \emph{noncommutative} Permanent
function does not have polynomial-size noncommutative circuits.

The connection that we show here between derandomizing the isolation
lemma and noncommutative circuit size lower bounds is based on the
above observation and our noncommutative polynomial identity test
based on the isolation lemma.

\subsection*{Commutative circuits}

Klivans and Spielman \cite{KS01} apply a more general form of the
isolation lemma to obtain a polynomial identity test (in the
commutative) case. This lemma is stated below.

\begin{lemma}\label{linearform}{\rm\cite[Lemma 4]{KS01}}
  Let $L$ be any collection of linear forms over variables
  $z_1,z_2,\cdots,z_n$ with integer coefficients in the range
  $\{0,1,\cdots,K\}$. If each $z_i$ is picked independently and
  uniformly at random from $\{0,1,\cdots,2Kn\}$ then with probability
  at least $1/2$ there is a unique linear form from $C$ that attains
  minimum value at $(z_1,\cdots,z_n)$.
\end{lemma}

We can formulate a restricted version of this lemma similar to
Lemma~\ref{mn-iso-lemma} that will apply only to sets of linear forms
$L$ accepted by a boolean circuit $C$. More precisely, an integer
vector $(\alpha_1,\cdots,\alpha_n)$ such that
$\alpha_i\in\{0,\cdots,K\}$ is in $L$ if and only if
$(\alpha_1,\cdots,\alpha_n)$ is accepted by the boolean circuit $C$.

Thus, for this form of the isolation lemma we can formulate another
derandomization hypothesis analogous to Hypothesis 1 as follows.\\

\noindent\textbf{Hypothesis 3.}~~There is a deterministic algorithm
$\mathcal{A}_3$ that takes as input $(C,n,K)$, where $C$ is a boolean
circuit that takes as input $(\alpha_1,\cdots,\alpha_n)$ such that
$\alpha_i\in\{0,\cdots,K\}$, and outputs a collection of weight
functions $w_1,w_2,\cdots,w_t$ such that $w_i:[n]\rightarrow [2Kn]$,
with the property that for some weight vector $w_i$ there is a unique
linear form $(\alpha_1,\cdots,\alpha_n)$ accepted by $C$ which attains
the minimum value $\sum_{j=1}^n w_i(j)\alpha_j$. Furthermore,
$\mathcal{A}_3$ runs in time subexponential in $\size(C)$.

\section{Automata Theory background}\label{automata-theory}

We recall some standard automata theory \cite{HU78}. Fix a finite
automaton $A = (Q,\delta,q_0,q_f)$ which takes inputs in $\{0,1\}^*$,
$Q$ is the set of states, $\delta:Q\times \{0,1\}\rightarrow Q$ is the
transition function, and $q_0$ and $q_f$ are the initial and final
states respectively (we only consider automata with unique accepting
states). For each $b\in\{0,1\}$, let $\delta_b:Q\rightarrow Q$ be
defined by: $\delta_b(q) = \delta(q,b)$. These functions generate a
submonoid of the monoid of all functions from $Q$ to $Q$. This is the
transition monoid of the automaton $A$ and is well-studied in automata
theory \cite[page 55]{Str94}. We now define the $0$-$1$ matrix $M_b
\in \F^{|Q|\times |Q|}$ as follows:
\[
M_b(q,q') = \left\{\begin{array}{cc}
1 & \textrm{if $\delta_b(q) = q'$,}\\
0 & \textrm{otherwise.} \end{array}\right.
\]

The matrix $M_b$ is the adjacency matrix of the graph of
$\delta_b$. As $M_b$ is a $0$-$1$ matrix, we can consider it as a
matrix over any field $\F$.

For a string $w=w_1w_2\cdots w_k\in\{0,1\}^*$ we define $M_w$ to be
the matrix product $M_{w_1}M_{w_2}\cdots M_{w_k}$. If $w$ is the empty
string, define $M_w$ to be the identity matrix of dimension $|Q|\times
|Q|$. Let $\delta_w$ denote the natural extension of the transition
function to $w$; if $w$ is the empty string, $\delta_w$ is simply the
identity function. We have
\begin{equation}
\label{dit_prelim_eqn}
M_w(q,q') = \left\{\begin{array}{cc}
1 & \textrm{if $\delta_w(q) = q'$,}\\
0 & \textrm{otherwise.} \end{array}\right.
\end{equation}
Thus, $M_w$ is also a matrix of zeros and ones for any string $w$.
Also, $M_w(q_0,q_f) = 1$ if and only if $w$ is accepted by the
automaton $A$.

\subsection{Noncommutative arithmetic circuits and automata}\label{dit_output}

This subsection is reproduced from \cite{AMS08} to make this paper
self-contained.

Consider the ring $\F\{x_1,\cdots,x_n\}$ of polynomials with
noncommuting variables $x_1,\cdots,x_n$ over a field $\F$. Let $C$ be
a noncommutative arithmetic circuit computing a polynomial $f\in
\F\{x_1,\cdots,x_n\}$. Let $d$ be an upper bound on the degree of
$f$. We can consider monomials over $x_1,\cdots,x_n$ as strings over
an alphabet of size $n$. For our construction, it is more convenient
to encode each $x_i$ as a string over $\{0,1\}$. We encode the
variable $x_i$ by the string $v_i=01^i0$. Clearly, each monomial over
the $x_i$'s of degree at most $d$ maps uniquely to a binary string of
length at most $d(n+2)$.

Let $A = (Q,\delta,q_0,q_f)$ be a finite automaton over the alphabet
$\{0,1\}$. We have matrices $M_{v_i}\in \F^{|Q|\times |Q|}$ as defined
in Section \ref{automata-theory}, where $v_i$ is the binary string
that encodes $x_i$. We are interested in the output matrix obtained
when the inputs $x_i$ to the circuit $C$ are replaced by the matrices
$M_{v_i}$. This output matrix is defined in the obvious way: the
inputs are $|Q|\times |Q|$ matrices and we do matrix addition and
matrix multiplication at each addition gate (respectively,
multiplication gate) of the circuit $C$. We define the \emph{output of
  $C$ on the automaton $A$} to be this output matrix $M_{out}$.
Clearly, given circuit $C$ and automaton $A$, the matrix $M_{out}$ can
be computed in time $\poly(|C|,|A|,n)$.

We observe the following property: the matrix output $M_{out}$ of $C$
on $A$ is determined completely by the polynomial $f$ computed by $C$;
the structure of the circuit $C$ is otherwise irrelevant. This is
important for us, since we are only interested in $f$. In particular,
the output is always $0$ when $f\equiv 0$.

More specifically, consider what happens when $C$ computes a
polynomial with a single term, say $f(x_1,\cdots,x_n) = cx_{j_1}\cdots
x_{j_k}$, with a non-zero coefficient $c\in\F$. In this case, the
output matrix $M_{out}$ is clearly the matrix $cM_{v_{j_1}}\cdots
M_{v_{j_k}} = cM_w$, where $w=v_{j_1}\cdots v_{j_k}$ is the binary
string representing the monomial $x_{j_1}\cdots x_{j_k}$. Thus, by
Equation \ref{dit_prelim_eqn} above, we see that the entry
$M_{out}(q_0,q_f)$ is $0$ when $A$ rejects $w$, and $c$ when $A$
accepts $w$. In general, suppose $C$ computes a polynomial $f =
\sum_{i=1}^{t}c_i m_i$ with $t$ nonzero terms, where $c_i\in
\F\setminus\{0\}$ and $m_i = \prod_{j=1}^{d_i}x_{i_j}$, where $d_i\leq
d$. Let $w_i = v_{i_1}\cdots v_{i_{d_i}}$ denote the binary string
representing monomial $m_i$.  Finally, let $S_A^f =
\{i\in\{1,\cdots,t\}~|~A \textrm{ accepts } w_i\}$.

\begin{theorem}\label{autothm}{\rm{\cite{AMS08}}}
Given any arithmetic circuit $C$ computing polynomial $f\in
\F\{x_1,\cdots,x_n\}$ and any finite automaton $A =
(Q,\delta,q_0,q_f)$, then the output $M_{out}$ of $C$ on $A$ is such
that $M_{out}(q_0,q_f) = \sum_{i\in S_A^f}c_i$.
\end{theorem}

\begin{proof}
  The proof is an easy consequence of the definitions and the
  properties of the matrices $M_w$ stated in Section \ref{automata-theory}.
  Note that $M_{out} = f(M_{v_1},\cdots,M_{v_n})$. But 
  $f(M_{v_1},\cdots,M_{v_n}) = \sum_{i=1}^{s}c_i M_{w_i}$, where $w_i
  = v_{i_1}\cdots v_{i_{d_i}}$ is the binary string representing
  monomial $m_i$. By Equation \ref{dit_prelim_eqn}, we know that
  $M_{w_i}(q_0,q_f)$ is $1$ if $w_i$ is accepted by $A$, and $0$
  otherwise. Adding up, we obtain the result.
\end{proof}

We now explain the role of the automaton $A$ in testing if the
polynomial $f$ computed by $C$ is identically zero. Our basic idea is
to design an automaton $A$ that accepts exactly one word from among
all the words that correspond to the nonzero terms in $f$. This would
ensure that $M_{out}(q_0,q_f)$ is the nonzero coefficient of the
monomial filtered out. More precisely, we will use the above theorem
primarily in the following form, which we state as a corollary.

\begin{corollary}{\rm{\cite{AMS08}}}
Given any arithmetic circuit $C$ computing polynomial $f\in
\F\{x_1,\cdots,x_n\}$ and any finite automaton $A =
(Q,\delta,q_0,q_f)$, then the output $M_{out}$ of $C$ on $A$ satisfies:
\label{autocor_use}
\begin{itemize}
\item[(1)] If $A$ rejects every string corresponding to a monomial in
$f$, then $M_{out}(q_0,q_f) = 0$.  
\item[(2)] If $A$ accepts exactly one string corresponding to a
monomial in $f$, then $M_{out}(q_0,q_f)$ is the nonzero coefficient of
that monomial in $f$.
\end{itemize}
Moreover, $M_{out}$ can be computed in time $\poly(|C|,|A|,n)$.
\end{corollary}

\begin{proof}
Both points ($1$) and ($2$) are immediate consequences of the above
theorem. The complexity of computing $M_{out}$ easily follows from its
definition.
\end{proof}

Another interesting corollary to the above theorem is the following.

\begin{corollary}\label{autocor_pick}{\rm{\cite{AMS08}}}
  Given any arithmetic circuit $C$ over $\F\{x_1,\cdots,x_n\}$, and
  any monomial $m$ of degree $d_m$, we can compute the coefficient of
  $m$ in $C$ in time $\poly(|C|,d_m,n)$.
\end{corollary}

\begin{proof}
  Apply Corollary \ref{autocor_use} with $A$ being any standard
  automaton that accepts the string corresponding to monomial $m$ and
  rejects every other string. Clearly, $A$ can be chosen so that $A$
  has a unique accepting state and $|A| = O(nd_m)$.
\end{proof}

\begin{remark}
  Corollary~\ref{autocor_pick} is very unlikely to hold in the
  commutative ring $\F[x_1,\cdots,x_n]$. For, it is easy to see that
  in the commutative case computing the coefficient of the monomial
  $\prod_{i=1}^nx_i$ in even a product of linear forms $\Pi_{i}\ell_i$
  is at least as hard as computing the permanent over $\F$, which is
  \mbox{$\#$\rm P}-complete when $\F=\mathbb{Q}$.
\end{remark}

\section{Noncommutative identity test based on isolation
  lemma}\label{main-result}

We now describe a new identity test for noncommutative circuits based
on the isolation lemma. It is directly based on the results from
\cite{AMS08}. This is conceptually quite different from the randomized
identity test of Bogdanov and Wee \cite{BW05}.

\begin{theorem}\label{ncomm-pit}
Let $f\in\F\{x_1,x_2,\cdots, x_n\}$ be a polynomial given by an arithmetic 
circuit $C$ of size $m$. Let $d$ be an upper bound on the degree of $f$. 
Then there is a randomized algorithm which runs in time $\poly(n,m,d)$ 
and can test whether $f\equiv 0$.  
\end{theorem}

\begin{proof}
  Let $[d]=\{1,2,\cdots,d\}$ and $[n]=\{1,2,\cdots,n\}$. Consider the
  set of tuples $U = [d]\times [n]$. Let $v=x_{i_1}x_{i_2}\cdots
  x_{i_t}$ be a nonzero monomial of $f$. Then the monomial can be
  identified with the following subset $S_v$ of $U$ :
\[
S_v = \{(1,i_1), (2,i_2), \cdots, (t,i_t)\}
\]  

Let $\mathcal{F}$ denotes the family of subsets of $U$ corresponding
to the nonzero monomials of $f$ i.e,
\[
\mathcal{F}=\{S_v ~|~v~\mbox{is a nonzero monomial in}~f\}
\]

By the Isolation Lemma we know that if we assign random weights from
$[2dn]$ to the elements of $U$, with probability at least $1/2$, there
is a unique minimum weight set in $\mathcal{F}$.  Our aim will be to
construct a family of small size automatons which are indexed by
weights $w\in [2nd^2]$ and $t\in[d]$, such that the automata $A_{w,t}$
will precisely accept all the strings (corresponding to the monomials)
$v$ of length $t$, such that the weight of $S_v$ is $w$.  Then from
the isolation lemma we will argue that the automata corresponding to
the minimum weight will precisely accept only one string
(monomial). Now for $w\in [2nd^2]$, and $t\in[d]$, we describe the
construction of the automaton $A_{w,t} = (Q, \Sigma, \delta, q_0, F)$
as follows: $Q = [d]\times [2nd^2]\cup \{(0,0)\}$, $\Sigma =
\{x_1,x_2,\cdots,x_n\}$, $q_0 = \{(0,0)\}$ and $F=\{(t,w)\}$.  We
define the transition function $\delta : Q\times\Sigma\rightarrow Q$,
\[
\delta((i,V),x_j) = (i+1, V+W),  
\] 
where $W$ is the random weight assign to $(i+1,j)$. Our automata family 
$\mathcal{A}$ is simply, 
\[
\mathcal{A}=\{A_{w,t}~|~w\in[2nd^2], t\in[d]\}.
\] 
Now for each of the automaton $A_{w,t}\in\mathcal{A}$, we mimic the
run of the automaton $A_{w,t}$ on the circuit $C$ as described in
Section~ \ref{automata-theory}. If the output matrix corresponding to
any of the automaton is nonzero, our algorithm declares $f\neq 0$,
otherwise declares $f\equiv 0$.

The correctness of the algorithm follows easily from the Isolation Lemma. 
By the Isolation Lemma we know, on random assignment, a unique set $S$ 
in $\mathcal{F}$ gets the minimum weight $w_{\min}$ with probability 
at least $1/2$. Let $S$ corresponds to the 
monomial $x_{i_1}x_{i_2}\cdots x_{i_{\ell}}$. Then the automaton $A_{w_{\min},\ell}$ 
accepts the string (monomial) $x_{i_1}x_{i_2}\cdots x_{i_{\ell}}$. Furthermore, 
as no other set in $\mathcal{F}$ get the same minimum weight, 
$A_{w_{min},\ell}$ rejects all the other monomials. So the $(q_0,q_f)$ entry 
of the output matrix $M_{o}$, that we get in running $A_{w_{\min},\ell}$ on 
$C$ is nonzero. Hence with probability at least $1/2$, our algorithm 
correctly decide that $f$ is nonzero. The success probability can be boosted 
to any constant by standard independent repetition of the same algorithm. 
Finally, it is trivial to see that the algorithm always decides correctly 
if $f\equiv 0$.  
\end{proof}

\section{Noncommutative identity testing and circuit lower
  bounds}\label{lowerbound}

For commutative circuits, Impagliazzo and Kabanets \cite{KI03} have
shown that derandomizing PIT implies circuit lower bounds. It implies
that either $\NEXP\not\subset \ppoly$ or the integer Permanent does
not have polynomial-size arithmetic circuits.

In \cite{AMS08} we have observed that this also holds in the
noncommutative setting.  I.e., if noncommutative PIT has a
deterministic polynomial-time algorithm then either
$\NEXP\not\subset \ppoly$ or the \emph{noncommutative} Permanent
function does not have polynomial-size noncommutative circuits.  We
note here that noncommutative circuit lower bounds are sometimes
easier to prove than for commutative circuits. E.g.\ Nisan \cite{N91} has
shown exponential-size lower bounds for noncommutative formula size
and further results are known for pure noncommutative circuits
\cite{N91,RS05}. However, proving superpolynomial size lower bounds
for general noncommutative circuits computing the Permanent has
remained an open problem.

To keep this paper self contained, we briefly recall the discussion
from \cite{AMS08}.

The noncommutative Permanent function $\Perm(x_1,\cdots,x_n)\in
R\{x_1,\cdots,x_n\}$ is defined as
\[
\Perm(x_1,\cdots,x_n)=\sum_{\sigma\in S_n}\prod_{i=1}^n x_{i,\sigma(i)},
\]
where the coefficient ring $R$ is any commutative ring with unity.
Specifically, for the next theorem we choose $R=\mathbb{Q}$.

Let $\SUBEXP$ denote $\cap_{\epsilon>0}\DTIME(2^{n^\epsilon})$ and
$\NSUBEXP$ denote $\cap_{\epsilon>0}\NTIME(2^{n^\epsilon})$.

\begin{theorem}\label{lower-bound}{\rm{\cite{AMS08}}}
  If $\PIT$ for noncommutative circuits of polynomial degree
  $C(x_1,\cdots,x_n)\in\mathbb{Q}\{x_1,\cdots,x_n\}$ is in $\SUBEXP$,
  then either $\NEXP\not\subset \ppoly$ or the \emph{noncommutative}
  Permanent function does not have polynomial-size noncommutative
  circuits.
\end{theorem}

\begin{proof}
  Suppose $\NEXP\subset \ppoly$.  Then, by the main result of
  \cite{IKW02} we have $\NEXP=\MA$.  Furthermore, by Toda's theorem
  $\MA\subseteq\P^{\Perm_{\mathbb{Z}}}$, where the oracle computes the
  integer permanent. Now, assuming $\PIT$ for noncommutative circuits
  of polynomial degree is in deterministic polynomial-time we will
  show that the (noncommutative) Permanent function does not have
  polynomial-size noncommutative circuits.  Suppose to the contrary
  that it does have polynomial-size noncommutative circuits. Clearly,
  we can use it to compute the integer permanent as well. Furthermore,
  as in \cite{KI03} we notice that the noncommutative $n\times n$
  Permanent is also uniquely characterized by the identities
  $p_1(x)\equiv x$ and $p_i(X)=\sum_{j=1}^i x_{1j} p_{i-1}(X_j)$ for
  $1<i \leq n$, where $X$ is a matrix of $i^2$ noncommuting variables
  and $X_j$ is its $j$-th minor w.r.t.\ the first row. I.e.\ if
  arbitrary polynomials $p_i, 1\leq i\leq n$ satisfies these $n$
  identities over \emph{noncommuting} variables $x_{ij}, 1\leq i,j\leq
  n$ if and only if $p_i$ computes the $i\times i$ permanent of
  noncommuting variables. The rest of the proof is exactly as in
  Impagliazzo-Kabanets \cite{KI03}. We can easily describe an NP
  machine to simulate a $\P^{\Perm_{\Z}}$ computation. The NP machine
  guesses a polynomial-size noncommutative circuit for $\Perm$ on
  $m\times m$ matrices, where $m$ is a polynomial bound on the matrix
  size of the queries made. Then the NP verifies that the circuit
  computes the permanent by checking the $m$ \emph{noncommutative}
  identities it must satisfy. This can be done in $\SUBEXP$ by
  assumption. Finally, the NP machines uses the circuit to answer all
  the integer permanent queries. Putting it together, we get
  $\NEXP=\NSUBEXP$ which contradicts the nondeterministic time
  hierarchy theorem.
\end{proof}

\section{The Results}

We are now ready to prove our first result. Suppose the
derandomization Hypothesis 1 holds (as stated in the introduction):
i.e.\ suppose there is a deterministic algorithm $\mathcal{A}_1$ that
takes as input $(C,n)$ where $C$ is an $n$-input boolean circuit and
in subexponential time computes a set of weight functions
$w_1,w_2,\cdots,w_t$, $w_i:[n]\rightarrow [2n]$ such that the set
system $\mathcal{F}_C$ defined by the circuit $C$ has a unique minimum
weight set w.r.t.\ at least one of the weight functions $w_i$.

Let $C'(x_1,x_2,\cdots,x_n)$ be a noncommutative arithmetic circuit of
degree $d$ bounded by a polynomial in $\size(C')$. By
Corollary~\ref{autocor_pick}, there is a deterministic polynomial-time
algorithm that takes as input $C'$ and a monomial $m$ of degree at
most $d$ and accepts if and only if the monomial $m$ has nonzero
coefficient in the polynomial computed by $C'$. Thus, we have a
boolean circuit $C$ of size polynomial in $\size(C')$ that accepts
only the (binary encodings of) monomials $x_{i_1}x_{i_2}\cdots
x_{i_k}$, $k\leq d$ that have nonzero coefficients in the polynomial
computed by $C'$. Now, as a consequence of Theorem~\ref{ncomm-pit} and
its proof we have a \emph{deterministic} subexponential algorithm for
checking if $C'\equiv 0$, assuming algorithm $\mathcal{A}_1$ exists.
Namely, we compute the boolean circuit $C$ from $C'$ in polynomial
time. Then, invoking algorithm $\mathcal{A}_1$ with $C$ as input we
compute at most subexponentially many weight functions
$w_1,\cdots,w_t$. Then, following the proof of Theorem~\ref{ncomm-pit}
we construct the automata corresponding to these weight functions and
evaluate $C'$ on the matrices that each of these automata define in
the prescribed manner. By assumption about algorithm $\mathcal{A}_1$,
if $C'\not\equiv 0$ then one of these $w_i$ will give matrix inputs
for the variables $x_j, 1\leq j\leq n$ on which $C'$ evaluates to a
nonzero matrix. We can now show the following theorem.

\begin{theorem}\label{result1}
  If the subexponential time algorithm $\mathcal{A}_1$ satisfying
  Hypothesis 1 exists then noncommutative identity testing is in
  $\SUBEXP$ which implies that either $\NEXP\not\subset \P/\poly$ or
  the Permanent does not have polynomial size noncommutative circuits.
\end{theorem}

\begin{proof}
The result is a direct consequence of the discussion preceding
the theorem statement and Theorem~\ref{lower-bound}.
\end{proof}

We now turn to the result under the \emph{stronger} derandomization
Hypothesis 2 (stated in the introduction). More precisely, suppose
there is a deterministic algorithm $\mathcal{A}_2$ that takes as input
$(m,n)$ and in time polynomial in $m$ computes a set of weight
functions $w_1,w_2,\cdots,w_t$, $w_i:[n]\rightarrow [2n]$ such that
for \emph{each} $n$-input boolean circuit $C$ of size $m$, the set
system $\mathcal{F}_C$ defined by the circuit $C$ has a unique minimum
weight set w.r.t.\ at least one of the weight functions $w_i$. We show
that there is an \emph{explicit} polynomial\footnote{By explicit we
  mean that the coefficients of $f$ are computable in time exponential
  in $n$.} $f(x_1,\cdots,x_n)$ in noncommuting variables $x_i$ that
does not have subexponential size noncommutative circuits.

\begin{theorem}\label{result2}
  Suppose there is a polynomial-time algorithm $\mathcal{A}_2$
  satisfying Hypothesis 2. Then for all but finitely many $n$ there is
  an explicit polynomial
  $f(x_1,\cdots,x_n)\in\F\{x_1,x_2,\cdots,x_n\}$ (where the field $\F$
  is either rationals or any finite field) in noncommuting variables
  $x_i$ that is computable in $2^{n^{O(1)}}$ time (by a uniform
  algorithm) and does not have noncommutative arithmetic circuits of
  size $2^{o(n\lg n)}$.
\end{theorem}


\begin{proof}
  Let $T_n$ denote the set of all sequences $(i_1,i_2,\cdots,i_n)$,
  for $i_j\in[n]$, $1\leq j\leq n$. For each such sequence
  $\alpha=(i_1,i_2,\cdots,i_n)\in T_n$ let $m_\alpha$ denote the
  monomial $x_{i_1}x_{i_2}\cdots x_{i_n}$. Now, we write
\[
 f(x_1,x_2,\cdots,x_n)=\sum_{\alpha\in T_n} c_\alpha m_\alpha,
\]
where we will pick the scalars $c_\alpha$ appropriately so that the
polynomial $f$ has the claimed property. Suppose $\mathcal{A}_2$ runs
in time $m^c$ for constant $c>0$, where $m$ denotes the size bound of
the boolean circuit $C$ defining set system $\mathcal{F}_C$. Notice
that the number $t$ of weight functions is bounded by $m^c$. As
explained in Theorem~\ref{ncomm-pit}, each weight function will give
rise to a collection of $2n^4$ automata $\A_k$, each of which will
prescribe matrices of dimension at most $r=\poly(n)$ to be assigned
for the input variables $x_j, 1\leq j\leq n$. Call these matrices
$M^{(k)}_{i,j}$. For each weight function $w_i$ write down linear
equations for each $k\in[2n^4]$.
\[
f(M^{(k)}_{i,1},M^{(k)}_{i,2},\cdots,M^{(k)}_{i,n})=0.
\]

This will actually give us a system of at most $2n^4r^2$ linear
equations in the unknown scalars $c_\alpha$. Since there are $t\leq
m^c$ weight functions in all, all the linear constraints put together
give us a system of at most $2n^4r^2m^c$ linear equations. Now, the
number of distinct (noncommuting) monomials $m_\alpha$ is $n^n=2^{n\lg
  n}$ which asymptotically exceeds $2n^4r^2m^c$ for $m=2^{o(n\lg n)}$,
since $r$ is polynomially bounded.  Thus, the system of linear
equations has a nontrivial solution in the $c_\alpha$'s that can be
computed using Gaussian elimination in time exponential in $n$.

Notice that the polynomial $f(x_1,\cdots,x_n)$, defined by the
solution to the $c_\alpha$'s, is a nonzero polynomial. We claim that
$f$ cannot have a noncommutative circuit of size $2^{o(n\lg n)}$.
Assume to the contrary that $C'(x_1,\cdots,x_n)$ is a noncommutative
circuit of size $s=2^{o(n\lg n)}$ for $f$. Then, by Corollary
~\ref{autocor_pick} there is an $n'$-input boolean circuit $C$ of size
$m=s^{O(1)}=2^{o(n\lg n)}$ that accepts precisely the (binary
encodings) of those monomials that are nonzero in $C'$. Let
$w_1,\cdots,w_t$ be the weight functions output by $\A_2$ for input
$(m,n')$. By Hypothesis 2, for some weight function $w_i$ and some
$k\in[2n^4]$ the circuit $C'$ must be nonzero on matrices
$M^{(k)}_{i,j}$. However, $f$ evaluates to zero, by construction, on
the matrix inputs prescribed by all the weight functions
$w_1,\cdots,w_t$. This is a contradiction to the assumption and it
completes the proof.
\end{proof}

\begin{remark}
  We can formulate both Hypothesis 1 and Hypothesis 2 more generally
  by letting the running time of algorithms $\mathcal{A}_1$ and
  $\mathcal{A}_2$ be a function $t(m,n)$. We will then obtain suitably
  quantified circuit lower bound results as consequence.
\end{remark}

\subsection*{Commutative circuits}


We now show that under the derandomization Hypothesis 3 (stated in the
introduction) we can obtain a stronger consequence than
Theorem~\ref{result1}.

\begin{theorem}\label{result3}
  If a subexponential-time algorithm $\mathcal{A}_3$ satisfying
  Hypothesis 3 exists then identity testing over $\Q$ is in $\SUBEXP$
  which implies that either $\NEXP\not\subset \P/\poly$ or the integer
  Permanent does not have polynomial size arithmetic circuits.
\end{theorem}

\begin{proof}
  Using Lemma~\ref{linearform} it is shown in \cite[Theorem 5]{KS01}
  that there is a randomized identity test for small degree
  polynomials in $\Q[x_1,\cdots,x_n]$, where the polynomial is given
  by an arithmetic circuit $\hat{C}$ of polynomially bounded degree
  $d$. The idea is to pick a random weight vector $w:[n]\rightarrow
  [2nd]$ and replace the indeterminate $x_i$ by $y^{w(i)}$, where $d$
  is the total degree of the input polynomial.  As the circuit
  $\hat{C}$ has small degree, after this univariate substitution the
  circuit can be evaluated in deterministic polynomial time to
  explicitly find the polynomial in $y$. By Lemma~\ref{linearform} it
  will be nonzero with probability $1/2$ if $\hat{C}$ computes a
  nonzero polynomial.

  Coming to the proof of this theorem, if $\NEXP\not\subset\P/\poly$
  then we are done. So, suppose $\NEXP\subset\P/\poly$. Notice that
  given any monomial $x_1^{d_1}\cdots x_n^{d_n}$ of total degree
  bounded by $d$ we can test if it is a nonzero monomial of $\hat{C}$
  in exponential time ( explicitly listing down the monomials of the
  polynomial computed by $\hat{C}$). Therefore, since
  $\NEXP\subset\P/\poly$ there is a polynomial-size boolean circuit
  $C$ that accepts the vector $(d_1,\cdots,d_n)$ iff $x_1^{d_1}\cdots
  x_n^{d_n}$ is a nonzero monomial in the given polynomial (as
  required for application of Hypothesis 3).

  Now, we invoke the derandomization Hypothesis 3. We can apply the
  Klivans-Spielman polynomial identity test, explained above, to the
  arithmetic circuit $\hat{C}$ for each of the $t$ weight vectors
  $w_1,\cdots,w_t$ generated by algorithm $\mathcal{A}_3$ to obtain a
  subexponential deterministic identity test for the circuit $\hat{C}$
  by the properties of $\mathcal{A}_3$. Now, following the argument of
  Impagliazzo-Kabanets \cite{KI03} it is easy to derive that the
  integer Permanent does not have polynomial size arithmetic circuits.
\end{proof}

\begin{remark}
  We formulate a stronger version of Hypothesis 3 to obtain a
  conclusion similar to Theorem~\ref{result2} for commutative
  circuits. For example we can formulate the hypothesis:

  There is a deterministic algorithm $\mathcal{A}_4$ that takes as
  input $(m,n,K)$ and outputs a collection of weight functions
  $w_1,w_2,\cdots,w_t$ such that $w_i:[n]\rightarrow [2n]$, with the
  property that for each size $m$, $n$-input \emph{oracle} boolean
  circuit $C^A$ (where $A$ is EXP-complete) that takes as input
  $(\alpha_1,\cdots,\alpha_n)$ such that $\alpha_i\in\{0,\cdots,K\}$,
  there is some weight vector $w_i$ for which there is a unique linear
  form $(\alpha_1,\cdots,\alpha_n)$ accepted by $C^A$ which attains
  the minimum value $\sum_{j=1}^n w_i(j)\alpha_j$. Furthermore,
  $\mathcal{A}_4$ runs in time polynomial in $m$.

  It is easy to see that, similar to Theorem~\ref{result2}, as a
  consequence of this hypothesis there is some explicit polynomial
  $f(x_1,\cdots,x_n)$ (i.e.\ computable in EXP) which does not have
  commutative circuits of subexponential size.
\end{remark}

\section{Discussion}

An interesting open question is whether derandomizing similar
restricted versions of the Valiant-Vazirani lemma also implies circuit
lower bounds. We recall the Valiant-Vazirani lemma as stated in the
original paper \cite{VV}.

\begin{lemma}\label{vv}
  Let $S\subseteq \{0,1\}^t$. Suppose $w_i, 1\leq i\leq t$ are picked
  uniformly at random from $\{0,1\}^t$. For each $i$, let $S_i=\{v\in
  S\mid v.w_j=0, 1\leq j\leq i\}$ and let $p_t(S)$ be the probability
  that $|S_i|=1$ for some $i$. Then $p_t(S)\geq 1/4$.
\end{lemma}

Analogous to our discussion in Section~\ref{isolation}, here too we
can consider the restricted version where we consider $S_C\subseteq
\{0,1\}^n$ to be the set of $n$-bit vectors accepted by a boolean
circuit $C$ of size $m$. We can similarly formulate derandomization
hypotheses similar to Hypotheses 1 and 2.

We do not know if there is another randomized polynomial identity test
for noncommutative arithmetic circuits based on the Valiant-Vazirani
lemma. The automata-theoretic technique of Section~\ref{main-result}
does not appear to work. Specifically, given a matrix
$h:\mathbb{F}_2^n\rightarrow \mathbb{F}_2^k$, there is no
deterministic finite automaton of size
$\poly(n,k)$ that accepts $x\in\mathbb{F}_2^n$ if and only if $h(x)=0$.\\

\noindent{\bf Acknowledgements.}~~We are grateful to Manindra Agrawal
for interesting discussions and his suggestion that
Theorem~\ref{result2} can be obtained from the stronger hypothesis. We
also thank Srikanth Srinivasan for discussions.

\end{document}